\documentstyle[prd,aps,eqsecnum,preprint,psfig,floats]{revtex}

\def\b{\beta}

\def\e{\epsilon}
\def\g{\gamma}
\def\d{\delta}
\def\p{\phi}
\def\vp{\varphi}
\def\vpt{\widetilde{\varphi}}
\def\s{\sigma}
\def\l{\lambda}

\def\del{\partial}
\def\ha{\frac{1}{2}}

\def\be{\begin{equation}}
\def\ee{\end{equation}}
\def\bea{\begin{eqnarray}}
\def\eea{\end{eqnarray}}
\def\psit{\widetilde{\psi}}
\begin{document}
\draft
\tighten

\preprint{\vbox{\hfill OHSTPY--HEP--T--96--011 \\
          \vbox{\vskip0.5in}
          }}

\title{The Vacuum in Light-Cone Field Theory}

\author{David G. Robertson}
\address{Department of Physics, The Ohio State University, Columbus,
OH 43210}

\date{March 14, 1996}

\maketitle

\begin{abstract}
This is an overview of the problem of the vacuum in light-cone field
theory, stressing its close connection to other puzzles regarding
light-cone quantization.  I explain the sense in which the light-cone
vacuum is ``trivial,'' and describe a way of setting up a quantum
field theory on null planes so that it is equivalent to the usual
equal-time formulation.  This construction is quite helpful in
resolving the puzzling aspects of the light-cone formalism.  It
furthermore allows the extraction of effective Hamiltonians that
incorporate vacuum physics, but that act in a Hilbert space in which
the vacuum state is simple.  The discussion is fairly informal, and
focuses mainly on the conceptual issues.  Additional technical details
of the construction described here will appear in a forthcoming paper
written with Kent Hornbostel.

\vspace{.3in}
\noindent Talk presented at {\sc Orbis Scientiae 1996}, Miami Beach,
FL, January 25--28, 1996.  To appear in the proceedings.
\end{abstract}

\pacs{ }

\section{INTRODUCTION}

The most striking aspect of light-cone field theories is surely the
claim that the vacuum state is simple, or even trivial.  The basic
argument for this is elementary.  We begin by noting that the
longitudinal momentum $P^+=P^0+P^3$ is positive semidefinite and
conserved in interactions.  However, from the free-particle dispersion
relation in light-cone coordinates,
\be
P^-={P_\perp^2+m^2\over P^+}\; ,
\label{disprel}
\ee
it follows that the only finite-energy states other than the bare
vacuum that have $P^+=0$ contain massless quanta with $P_\perp=0$---a
set of measure zero in more than 1+1 dimensions.  Thus if these can be
ignored as unimportant in the full theory, then the bare vacuum is the
only state in the theory with $P^+=0$, and so must be an exact
eigenstate of the full interacting Hamiltonian.

If true, this enormously simplifies any effort to solve field theories
nonperturbatively using Hamiltonian diagonalization, as attempts to
compute the spectrum and wavefunctions of physical states are not
complicated by the need to recreate a ground state in which processes
occur at unrelated locations and energy scales.  It furthermore
results in a constituent picture, in which all quanta in, say, a
hadron's wavefunction are specifically related to that hadron.  This
allows for an unambiguous definition of the partonic content of
hadrons, and makes interpretation of the wavefunctions
straightforward.  For this reason the light-cone framework is the most
natural one in which to encode hadronic properties \cite{1}.

It immediately raises the question, however, of whether field theories
quantized on the light cone are equivalent in all respects with the
corresponding equal-time theories.  In many cases, notably QCD, there
is important physics that in the usual language is attributed to the
properties of the vacuum.  The fact that the pion is light on the
hadronic scale is conventionally traced to spontaneous breaking of the
axial flavor symmetries.  It is not clear how a trivial vacuum can be
a spontaneously broken vacuum of anything.  Furthermore, the QCD
ground state is a superposition of ``topological'' vacua labeled by a
parameter $\theta$.  This structure is important for understanding why
there is no ninth light pseudoscalar meson (for three light flavors),
and for satisfying cluster decomposition.  In addition, if
$\theta\neq0$ (and none of the quarks are massless), then QCD violates
$P$ and $T$, leading for example to a nonzero electric dipole moment
for the neutron.  The structure of the vacuum in QCD is clearly
connected to observable consequences of the theory.

One way of reconciling the apparent triviality of the vacuum in
light-cone quantization with the fact that the vacuum has nontrivial
physical consequences is based on the following set of observations
\cite{2}.  Eqn. (\ref{disprel}) implies that particle states that can
mix with the bare vacuum are high-energy states.  This suggests that
it is natural to think in terms of effective Hamiltonians.  That is,
we may imagine introducing an explicit cutoff on longitudinal momentum
for particles
\be
P^+>\l\; .
\ee
(For practical calculations it may be preferable to implement the
cutoff in some more sophisticated way.  I use this simple scheme to
illustrate the point.)  This immediately makes the vacuum trivial and
gives the resulting constituent picture.  In principle, it should then
be possible to restore the effects of states with $P^+<\l,$ including
the physics of the vacuum, by means of effective interactions which,
because the states eliminated have large light-cone energies, will be
local in $x^+$.\footnote{It is instructive to contrast this with the
analogous procedure in equal-time quantization.  In this case there
are states that are kinematically allowed to mix with the vacuum at
all energy scales, so that a description of vacuum physics in terms of
an effective Hamiltonian is not practicable.}  From this point of
view, the statement of the vacuum problem on the light cone is: How
can we compute these effective interactions?

I should perhaps emphasize at this point that the physical vacuum
state in light-cone field theory is not really trivial.  It cannot be,
if the light-cone theory is to be equivalent to the corresponding
equal-time theory.  The potential advantage of the light-cone approach
is that it may allow us to move the effects of the vacuum into the
Hamiltonian and work with a simple vacuum state.  The problem is that
sorting out the structure of the vacuum in light-cone quantization is
especially difficult.  At the kinematical point at which the vacuum
lives there is a diverging density of states with diverging energies,
but for which matrix elements of the Hamiltonian also are singular.
This is the loophole in the simple argument for vacuum triviality:
states with $P^+=0$ {\em do} have divergent energies, but they are
infinitely strongly coupled to low-energy states.  It therefore does
not follow that they decouple from low-energy physics.

Some sort of small-$P^+$ regularization is necessary even without the
motivation of obtaining a constituent picture, and how we approach the
problem of computing the effective interactions will certainly depend
on the specific choice of this regulator.  For example, if we think in
terms of simple momentum cutoffs then a renormalization group approach
is natural \cite{3} (the effective interactions are the counterterms
that remove dependence on the cutoff $\l$ from physical quantities).
Another method that has been studied in some detail involves using
discretization as an infrared regulator \cite{4}.  In this case any
vacuum structure must be connected with the properties of the zero
modes (in the Fourier sense) of the fields.

Both of these cutoffs, however, force the vacuum to be trivial from
the outset, and we face the problem of how to put the vacuum physics
back in.  The RG approach should work in principle---for a particular
set of cutoffs there is presumably a single effective Hamiltonian that
gives completely cutoff-independent results and corresponds to the QCD
Hamiltonian.  The difficulty here is a practical one.  It is not clear
whether this program can really be carried out, particularly since the
use of perturbation theory to construct the RG is probably not
sufficient for QCD.  The development of nonperturbative RGs for use in
light-cone field theory is an extremely interesting and challenging
problem.

Discretized light-cone quantization (DLCQ) is a self-contained
formalism with the inclusion of the zero modes, so the question here
is simply whether the vacuum physics is present or not at the end of
the day.  Analysis of simple models suggests that certain mean-field
aspects of the vacuum are in fact captured by the zero modes \cite{5};
however, not all of the relevant structure is present.  In the
$\p^4_{\rm 1+1}$ model, for example, the zero mode gives rise to the
expected strong-coupling phase transition, but with the (incorrect)
mean-field critical exponent.  This suggests that even with zero modes
DLCQ does not contain all of the necessary physics.

The main purpose of this talk is to describe one way of regulating
light-cone field theories that does not force the vacuum to be trivial
from the beginning.  In fact, the construction is designed to be
completely equivalent to the usual equal-time formalism, so that there
is no confusion about what is or is not present in the theory.  I
shall also sketch how it can be used to extract effective light-cone
Hamiltonians, which incorporate the physical effects of the vacuum but
act in a space with a trivial ground state.

In order to widen the context of the discussion, however, let me
mention a few other ``classic'' puzzles regarding light-cone
quantization.  All of these problems, including that of the vacuum,
are interrelated, and all of them are clarified by the construction I
shall describe.

\begin{itemize}
\item{}
It has been noted on occasion that a null plane is not a good Cauchy
surface for a relativistic wave equation; in fact, it is a surface of
characteristics for such an equation \cite{6}.  Thus initial
conditions on one such surface plus a Hamiltonian are not in general
sufficient to determine a general solution to the field equations.  In
the quantum field theory this problem is manifested as missing degrees
of freedom.  A particularly clear example of this is that of a free
massless fermion in two spacetime dimensions, as first discussed by
McCartor \cite{7}.  Here we find, following the usual light-cone
procedure, that the theory contains only right-moving particles; half
of the solution space is missing.  Additional degrees of freedom,
initialized on a second null plane, are required to properly
incorporate the left-movers.  One may have intuitively the idea that
this is a problem specific to massless fields in 1+1 dimensions (or
for only those massless particles with $p_\perp=0$ in higher
dimensions---a set of measure zero), but I shall argue that this is
not the case.

\item{} 
It is not clear what, if any, boundary conditions can be consistently
imposed on a null plane.  The problem is that some points on the
surface are causally connected, and requiring fields at such points to
be related in some way may result in an over-determined system.  In
other words, some boundary conditions may be in conflict with the
dynamics we wish to solve.

\item{} 
There seems to be a mysterious discontinuity in the way massless and
massive fields are treated on the light cone.  As mentioned above, for
a free massless fermion in 1+1 dimensions we must include independent
degrees of freedom on two different null planes to recover the correct
theory.  For nonzero mass, however, this does not seem to be
necessary; the usual light-cone construction gives a theory that is
precisely isomorphic to the equal-time theory of a free massive Fermi
field.  Thus the limit is apparently not smooth---it seems that
additional degrees of freedom must be switched on precisely at $m=0$.
We shall argue below that this apparent discontinuity is an artifact
of considering only {\em free} field theories.  In an interacting
theory a small-$k^+$ regulator must be employed, and this allows for a
completely unified treatment of massive and massless fields.  In
particular, the massless limit may be taken smoothly.

\item{}
Finally let us mention the ``No-problem'' problem.  This is that
simply ignoring most or all of these problems sometimes works.  For
example, the spectrum of the Schwinger model can be computed rather
well without worrying about any of the above \cite{8}.  This is
somewhat astonishing considering that this is a situation where these
problems might be expected to be at their most severe.  All the fields
are massless, the free fermion theory is known to be missing half of
the necessary states, and all points on the initial-value surface are
in causal contact.  There are many aspects of the Schwinger model that
simply cannot be understood without worrying about these
problems---the anomaly, the occurrence of a condensate, the
$\theta$-vacuum \cite{9}---but the spectrum of physical bosons is
reproduced exactly.  It would be nice if this feature were to hold in
more complicated examples, that certain things would be computable
even without a complete formulation incorporating all the subtleties.
One of the goals of the work I will describe is to try to understand
this in a quantitative way, by starting from a manifestly complete
formalism so that the effects of discarding certain pieces of physics
can be studied without ambiguity.

\end{itemize}

\begin{figure}
\centerline{
\psfig{figure=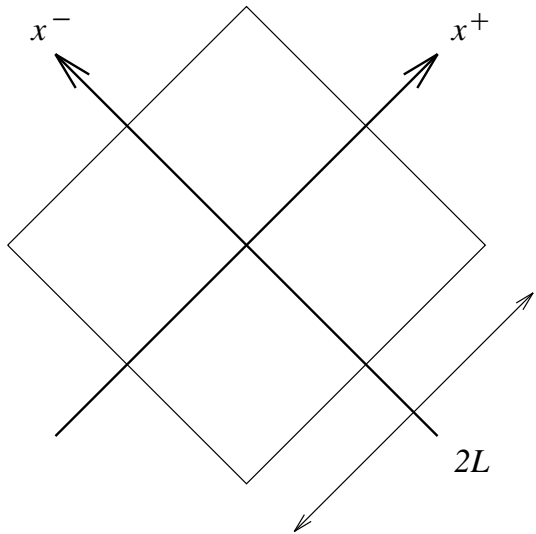}
}
\vspace{.1in}
\caption{Finite volume used for infrared regulation.}
\end{figure}

The rest of this paper is organized as follows.  In the next section I
discuss infrared regularization using a finite volume, and the
initial-value problem in such a box.  I show that with a careful
treatment of the boundary surfaces this formulation is completely
equivalent to equal-time quantization.  In particular, all conserved
charges (including the Hamiltonian) are guaranteed to be identical to
the operators we would construct in equal-time quantization, and the
vacuum state is complicated.  I then describe the formalism in detail
for scalar and Fermi fields.  Finally, I sketch how to use this
construction to obtain effective Hamiltonians for use in light-cone
field theories with a simple vacuum.

\pagebreak
\section{ LIGHT-CONE INITIAL VALUE PROBLEM IN A FINITE VOLUME}

As discussed above, one aspect of the vacuum problem on the light cone
is that it is difficult to find an infrared regulator that does not
automatically remove the vacuum structure.  One way of achieving this
involves using a finite volume as a regulator and treating the
initial-value problem carefully \cite{7,10,11}.  This analysis will
show that extra degrees of freedom, beyond what is present in the
conventional light-cone quantization, are required to construct a
theory that is equivalent to the equal-time theory.  These additional
degrees of freedom are what will allow complicated vacuum structure to
occur.

To be more concrete, let us consider a free massive scalar field in
1+1 dimensions confined to the region shown in Fig. 1.  (The extension
to higher dimensions is straightforward and is discussed in
\cite{11}.)  The first question is: What classical data are required
in order to determine a completely general solution to the field
equation
\be
\left(4\del_+\del_-+m^2\right)\p=0\; ?
\label{scalareom}
\ee
These data will correspond to independent operators in the quantum
field theory.

Let us begin by assuming that we have specified the value of $\p$ on
the conventional light-cone initial-value surface $x^+=0$, for $-L\leq
x^-\leq L$.  Eqn. (\ref{scalareom}) then allows us to evolve
$\del_-\p$ infinitesimally in $x^+$:
\be
\del_-\p(\e, x^-) = \del_-\p(0,x^-) + \e \del_+(\del_-\p(0,x^-))\; ,
\ee
where
\be
\del_+\left(\del_-\p\right)=-{m^2\over4}\p\; .
\ee
In order to go on, however, we need $\p$ itself on the new
$x^+$-slice.  This is obtained by integration,
\be
\p(\e,x^-)=\int_{-L}^{x^-}dy^- \del_-\p(\e,y^-)+\p(\e,-L)\; ,
\ee
which of course requires knowing the value of the field at one point
on the surface; here we have arbitrarily taken this point to be
$(\e,-L)$.  Now that we have 
$\p$ at $x^+=\e$, we can continue to the next slice,
and so on.  Clearly the process can be iterated to fill out the entire
box, provided that in addition to knowing $\p$ on $x^+=0$, we are also
given $\p$ along a surface of constant $x^-$.

We thus find that in order to obtain a general solution to Eqn.
(\ref{scalareom}) we must specify $\p$ both at $x^+=0$ (or more
generally on any surface of constant $x^+$) and on a surface of
constant $x^-$.  This means that when we build the quantum field
theory we should include independent operators initialized on some
surface of constant $x^-$ that represent this classical freedom.  This
simple analysis does not tell us {\em which} surfaces of constant
$x^-$ are most useful, however.  In order to answer this question it
is helpful to think about actually setting up the quantum theory with
fields initialized on both surfaces.  The basic problem is to write
down expressions for the fields and find a set of commutation
relations among them such that, e.g., the Poincar\'e generators
satisfy the correct algebra and generate the proper transformations on
the fields.  One important point is that in all but the simplest cases
(e.g., free field theories), the commutation relations between field
operators at time-like separations are {\em a priori} unknown.  We
should therefore only consider surfaces that contain no time-like
separated points.  Two particularly useful surfaces of this type are
shown in Figs. 2(a) and (b).  The surface shown in Fig. 2(c) is not
suitable.

\begin{figure}
\centerline{
\psfig{figure=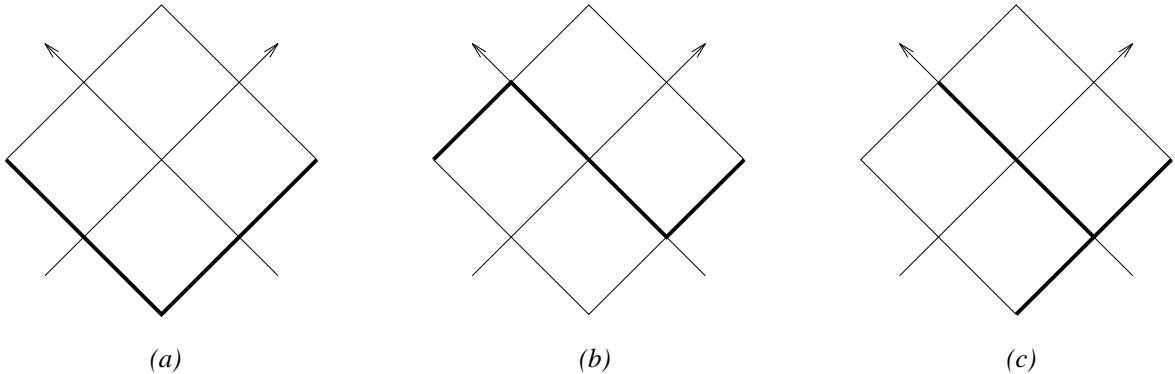,width=6.1in}
}
\vspace{.1in}
\caption{\noindent Some possible initial-value surfaces.  The dark
lines indicate where initial data are specified.}
\end{figure}

The two ``good'' surfaces each have particular advantages.  The
symmetric one [Fig. 2(a)] is manifestly parity invariant, for example.
(Recall that under parity $x^+\leftrightarrow x^-$, so that the two
wings are interchanged.)  This is helpful because initializing the
fields is completely symmetric on the two wings.  The asymmetric
surface [Fig. 2(b)] will be more useful, however, when we attempt to
extract an effective Hamiltonian for use in light-cone theories of the
usual type, that is, involving the fields initialized at $x^+=0$.

In fact, the fields on the asymmetric surface can be obtained from
those on the symmetric one by means of a shift in $x^+$ and a minor
reorganization of the degrees of freedom \cite{11}.  I shall therefore
concentrate first on the symmetric surface, where initialization is
easier, and shift to obtain the formulation on the asymmetric surface
only when we are ready to discuss effective Hamiltonians.

\subsection{Conserved Charges}

How do these additional degrees of freedom enter the dynamics?  That
is, how are they incorporated into the Hamiltonian and other conserved
charges of the theory?  As we shall see, understanding this leads to
certain clarifications regarding the equivalence of the light-cone and
equal-time formalisms.

The basic point is quite simple \cite{7}.  For any conserved current,
\be
\del_\mu J^\mu=0\; ,
\ee
we have
\be
\oint d\s_\mu J^\mu = 0\; ,
\label{cons}
\ee
where the integral is taken over a closed surface.  If this is taken
to be the surface shown in Fig. 3, for example, then Eqn. (\ref{cons})
implies
\bea
Q &=& \int_A^D dx J^0
\\
&=& \ha\int_A^B dx^+ J^- + \ha\int_C^B dx^- J^+ 
+ \ha\int_C^D dx^+ J^-\; .
\eea
(The factors of $1/2$ are due to our definition of light-cone
coordinates.)  The middle term in the second line is the usual
expression for the charge in light-cone quantization.  We therefore
see that inclusion of contributions from the boundaries is in general
{\em necessary} if the various charges are to be identical to those we
would construct in equal-time quantization.

\begin{figure}
\centerline{
\psfig{figure=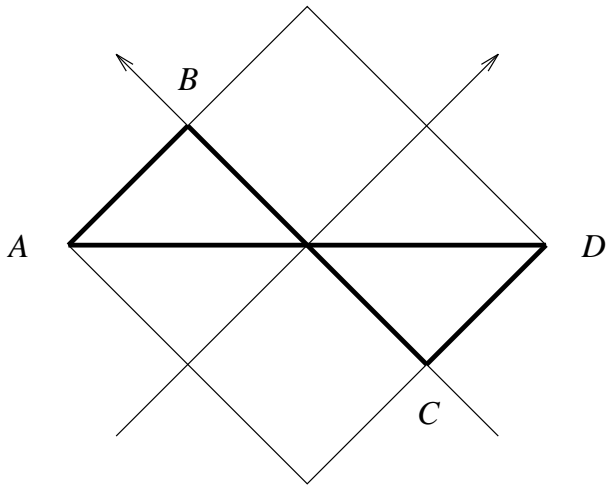}
}
\vspace{.1in}
\caption{Integration contour for conserved charges.}
\end{figure}

To see why the vacuum is complicated when the boundaries are retained,
consider the particular case of the energy-momentum tensor.  We have,
for example,
\be
P^+= \ha\int_A^B dx^+ T^{-+} + \ha\int_C^B dx^- T^{+-} 
+\ha\int_C^D dx^+ T^{-+}\; .
\ee
The boundary contributions involve $T^{-+}$, which contains
interaction terms.  Thus $P^+$ will no longer be diagonal in a Fock
basis, and the physical vacuum state can be complicated.

Recognizing the role played by the boundaries is crucial
in resolving certain specific puzzles of light-cone quantization.  For
example, in the Schwinger model (electrodynamics of massless fermions
in 1+1 dimensions) the vector and axial-vector currents are related by
\bea
J^+_5 &=& J^+
\\
J^-_5 &=& -J^-\; .
\eea
In the usual light-cone formalism, where charges are obtained by
integrating only over $x^+=0,$ this implies
\be
Q = \ha\int dx^- J^+ = Q_5\; .
\ee
But $Q_5$ is supposed to be anomalous, while $Q$ is conserved.  The
resolution of this paradox is that contributions from the boundary
surfaces also must be included.  These involve integrals of
$J^-_{(5)}$ over a surface of constant $x^-$, so that a difference
between $Q$ and $Q_5$ can arise \cite{9}.

It might be thought that this sort of problem arises only for massless
fields in 1+1 dimensions, or for the $p_\perp=0$ modes (only) of a
massless field in higher dimensions.  This is incorrect, however, and
is a result of considering only free field theories.  In a free
massive theory the limit of large box size can trivially be taken,
since there are no divergent matrix elements of the
Hamiltonian.\footnote{More precisely, there is no divergent coupling
between states.  The free energies diverge for $p^+\rightarrow0,$ but
we can argue that this is irrelevant, as the states for which this
occurs are not in the spectrum of either the light-cone or equal-time
theories---they have infinite energy.}  Thus the boundary fields
decouple completely and what remains is the usual light-cone theory.
This does not happen for a massless field (or for the $p_\perp=0$
modes of a higher-dimensional theory); in this case the left- and
right-moving modes are not coupled and so they can never decouple.
This is what leads to the apparent paradox regarding the difference
between massless and massive fields on the light cone, and is a point
I would like particularly to stress.  The need to include boundary
fields in order to obtain a light-cone formulation that is equivalent
to the equal-time theory is quite general, and applies to both massive
and massless fields in any number of dimensions.  What leads to
confusion is that in free field theory the boundaries can be decoupled
for massive fields.  (This is because the vacuum really {\em is}
trivial for free field theories!)  In an interacting theory, however,
the infinite-volume limit need not be trivial, and singularities in
the interactions can lead to observable consequences---vacuum
condensates, for example---that would simply be absent if the theory
were formulated without the boundary degrees of freedom.

In the next two sections I will show how to realize this scheme in
some simple cases.  The central problem is to obtain general
expressions for the fields on the initial surfaces such that all
requirements of relativistic quantum field theory, in particular
causality, are satisfied, and no boundary conditions are employed that
are in conflict with the dynamics.  The discussion will necessarily be
rather incomplete; my goal is mainly to give a flavor of how the
formalism is set up.  Further examples and many additional details
are given in \cite{11}.

\section{SCALARS}

Let us begin by considering a free, massless scalar field, quantized
on the symmetric surface I.\footnote{This theory does not actually
exist except with an infrared regulator such as the box in place.
The fact that massless elementary scalars are afflicted with
unmanageable infrared divergences is the essence of Coleman's theorem
regarding the impossibility of spontaneous breaking of continuous
symmetries in two spacetime dimensions; the necessary Goldstone boson
cannot exist.  Here we shall never try to remove the regulator, so
that the theory is always well defined.}  This is a very simple
example, but it serves to illustrate many of the basic points.

It is convenient to impose periodic boundary conditions at equal
time,
\be
\p(-L,L) = \p(L,-L)\; .  
\label{etper}
\ee
This insures that the Hamiltonian and other conserved charges are
time-independent.  Note that because the two points thus related are
separated by a space-like interval, there is no conflict with
causality.  The equation of motion is simply
\be 
\del_+\del_-\p = 0\; , 
\label{masslesseom}
\ee 
which has as its general solution 
\be 
\p(x^+,x^-) = \p_+(x^+)+\p_-(x^-)\; , 
\ee 
where $\p_\pm$ are arbitrary functions.  It is clear that initial data
on both $x^+=-L$ and $x^-=-L$ are required to obtain the full solution
space of the theory.  Our first problem is to find the most general
expression for the fields consistent with, e.g., causality, the
equations of motion, and the condition (\ref{etper}).

Now the equation of motion itself tells us what boundary conditions
are allowed on the light-cone surfaces.  To see this, integrate it
over one or the other of the wings.  We obtain, for example,
\be
\del_-\p(L,-L) = \del_-\p(-L,-L)\; ,
\ee
which implies, because of the equal-time periodicity condition,
\be
\del_-\p(-L,L) = \del_-\p(-L,-L)\; .
\ee
Thus the {\em derivative} $\del_-\p$ may be taken to be periodic on
$x^+=-L.$ Similarly, $\del_+\p$ may be taken to be periodic on the
surface $x^-=-L.$ We can therefore write general Fourier expansions
for these,
\begin{eqnarray}
\del_-\p(-L,x^-) & = & \del_-\vp(x^-)+{\p_0\over L}
\label{derivs1}\\
\nonumber \\
\del_+\p(x^+,-L) & = & \del_+\vpt(x^+)+{\widetilde{\p}_0\over L}\; ,
\label{derivs2}
\end{eqnarray}
where $\vp$ and $\vpt$ are sums of periodic oscillators and $\p_0$
and $\widetilde{\p}_0$ are the zero-momentum modes.  (The factors of
$L$ are introduced for later convenience.)  Explicitly,
\begin{eqnarray}
\vp(x^-) &=& {1\over\sqrt{2\pi}}\sum_{q=2,4,\dots} {1\over\sqrt{q}}
\left(a_q e^{-iq\pi x^-/2L} + a^\dagger_q e^{+iq\pi x^-/2L}\right)
\\
\nonumber \\
\vpt(x^+) &=& {1\over\sqrt{2\pi}}\sum_{q=2,4,\dots} {1\over\sqrt{q}}
\left(\widetilde{a}_q e^{-iq\pi x^+/2L} 
	+ \widetilde{a}^\dagger_q e^{+iq\pi x^+/2L}\right)\; .
\label{fourier}
\end{eqnarray}
We can then integrate Eqns. (\ref{derivs1})--(\ref{derivs2}) to obtain
the fields on the initial value surfaces, in terms of the given
derivatives and the value of the field at the corner point $(-L,-L)$.
We obtain
\begin{eqnarray}
\p(-L,x^-) &=& \vp(x^-)-\vp(-L)+{\p_0\over L}\left(x^-+L\right)
+\p(-L,-L)
\\ \nonumber \\
\p(x^+,-L) &=& \vpt(x^+)-\vpt(-L)+{\widetilde{\p}_0\over L}
\left(x^++L\right)+\p(-L,-L)\; .
\label{integrated}
\end{eqnarray}
Imposing the condition (\ref{etper}) then leads to the identification
\be
\p_0 = \widetilde{\p}_0\; .
\ee
Finally, it is convenient to define
\be
\p(-L,-L)\equiv \vp(-L) + \vpt(-L) +\psi
\ee
so that
\begin{eqnarray}
\p(-L,x^-) &=& \vp(x^-)+\vpt(L) + {\p_0\over L}\left(x^-+L\right)
+\psi
\\ \nonumber \\
\p(x^+,-L) &=& \vpt(x^+)+\vp(L)+{\p_0\over L}
\left(x^++L\right)+\psi\; .
\end{eqnarray}
These represent the most general expressions for the fields on the
initial surfaces that are consistent with the equations of motion and
the periodicity condition (\ref{etper}).  The classical initial data
are the Fourier modes $a_q$ and $\widetilde{a}_q,$ and the numbers
$\p_0$ and $\psi$.  In the quantum theory these become operators whose
commutation relations are to be determined by demanding that the
correct Heisenberg equations and Poincar\'e algebra are
obtained.\footnote{Note that the regulator breaks longitudinal boost
invariance, so that the commutator $[P^\pm,K]=\mp 2iP^\pm$ is not
recovered for finite $L$.  In practice the Heisenberg equations are
usually sufficient to determine the field algebra, perhaps
supplemented by the requirement that $[P^+,P^-]=0$ and that the fields
commute for space-like separations.}

Note that the fields are not themselves periodic on either of the
surfaces.  Imposing periodicity on the fields, as one would do in
DLCQ, corresponds to a less general expression for the fields, or,
equivalently, to focusing on a subset of the classical solution space.
In the quantum theory this corresponds to omitting degrees of freedom.

The Poincar\'e generators receive contributions from both wings of the
initial-value surface:
\be
P^\mu=\ha\int_{-L}^L dx^+ T^{-\mu}(x^+,-L) 
+ \ha\int_{-L}^L dx^- T^{+\mu}(-L,x^-)\; .
\label{pmusym}
\ee
The only nonvanishing components of the energy-momentum tensor are
\bea
T^{++} & = & 4(\del_-\p)^2\\
T^{--} & = & 4(\del_+\p)^2\; ,
\eea
so that
\begin{eqnarray}
P^+ &=& 2\int_{-L}^L dx^- \left[ \del_-\p(-L,x^-)\right]^2
\\
P^- &=& 2\int_{-L}^L dx^+ \left[ \del_+\p(x^+,-L)\right]^2\; .
\end{eqnarray}
These are guaranteed to be the same operators, though expressed in a
different representation, as we would obtain in an equal-time
quantization of this theory.  Requiring that the Heisenberg equations
reproduce the field equation (\ref{masslesseom}) leads to the
commutation relations
\begin{eqnarray}
\left[\vp(x^-),\vp(y^-)\right] &=& 
	-{i\over4}\e(x^--y^-)+{i\over4L}(x^--y^-)
\\
\left[\vpt(x^+),\vpt(y^+)\right] &= &
	-{i\over4}\e(x^+-y^+)+{i\over4L}(x^+-y^+)
\\
\left[\psi,\p_0\right] &=& {i\over4}\; ,
\label{psiphi}
\end{eqnarray}
where $\e$ is the antisymmetric step function.  In terms of the Fock
operators, the first two commutators correspond to
\be
\left[a_q,a^\dagger_p\right] =
\left[\widetilde{a}_q,\widetilde{a}^\dagger_p\right] = \d_{q,p}\; .
\ee
Furthermore, with the redefinitions
\bea
\psi & \equiv & {1\over2\sqrt{2}}(b+b^\dagger)\\
\phi_0 & \equiv & {i\over2\sqrt{2}}(b^\dagger-b)\; ,
\eea
Eqn. (\ref{psiphi}) can be recast as $[b,b^\dagger]=1$.

With these commutation relations the Poincar\'e generators correctly
translate the fields in $x^\pm,$ and furthermore satisfy,
$[P^+,P^-]=0$.  We thus have a satisfactory representation of the
theory, including all the relevant degrees of freedom.  The Fock space
is spanned by states of the form
\be
a^\dagger_{p_1}\dots a^\dagger_{p_n}
\widetilde{a}^\dagger_{q_1}\dots \widetilde{a}^\dagger_{q_m}
(b^\dagger)^l |0\rangle\; ,
\ee
where $|0\rangle$ is annihilated by $a_q$, $\widetilde{a}_q$, and $b$.
The Poincar\'e generators are given by
\bea
P^- &=& \sum_q\left({q\pi\over L}\right)
\widetilde{a}^\dagger_q \widetilde{a}_q+{4\over L}\left(\p_0\right)^2
\label{pp}
\\
P^+ &=& \sum_q\left({q\pi\over L}\right)
{a}^\dagger_q {a}_q + {4\over L} \left(\p_0\right)^2\; .
\label{pm}
\eea
In fact, the entire construction can be shown to be correct by mapping
it onto the equal-time theory of a massless scalar in a box.  That is,
we can quantize the theory at equal time with the periodicity
condition (\ref{etper}), and solve the resulting theory to obtain the
fields on the boundary surfaces.  We then identify the light-cone Fock
operators in terms of the equal-time ones, and so obtain the relation
between the two representations.  Not surprisingly, the $a_q$ modes
correspond to right-moving quanta, the $\widetilde{a}_q$ to
left-moving quanta, and $\p_0$ to the zero-momentum mode in the
equal-time Fourier expansion.  The Hamiltonian and momentum operators
can also be constructed and are identical to the appropriate
combinations of (\ref{pp}) and (\ref{pm}).

The massive case is more involved, essentially because the
left and right movers are coupled by the mass term (for example, a
finite boost can now change a left mover to a right mover, and vice
versa).  The commutator we wish to reproduce in this case is the
Pauli-Jordan function
\be
\left[\p(x),\p(y)\right]=-{i\over4}\left[\e(x^+-y^+)+\e(x^--y^-)\right]
J_0\left(\mu\sqrt{(x-y)^2}\right)\; , 
\ee
where $\mu$ is the mass.  This is properly causal; it vanishes
whenever $(x-y)$ is space-like.  Furthermore, the Bessel function is
unity on the light cone, reproducing the usual light-cone commutator
of the fields.  Light-cone commutators involving {\em derivatives} of
the field, however, can be $\mu$-dependent.  For example,
\be
\left[\del_-\p(-L,-L),\del_+\p(x^+,-L)\right]={i\mu^2\over16}\; .  
\ee
This suggests that $\mu$ should be built in to the expressions for the
fields, in such a way that those commutators involving derivatives
that are needed are correctly reproduced.

This can in fact be done, and the details are given in \cite{11}.  For
now, however, let us turn to the case of a massive Fermi field.  This
is similar in many ways due to the presence of the mass, but allows
for a slightly less cluttered treatment.

\section{FERMIONS}

Consider a free, massive Dirac fermion in 1+1 dimensions, quantizing
again on the symmetric surface.  The standard light-cone decomposition
of a Fermi field,
\be
\psi_\pm = \ha\g^0\g^\pm\psi\; ,
\ee
allows the separation of the Dirac equation into the coupled pair
\bea
i\del_+\psi_+ &=& {m\over2}\psi_-\\
\label{psip}
\nonumber \\
i\del_-\psi_- &=& {m\over2}\psi_+\; .
\label{psim}
\eea
A classical analysis of these, analogous to that given in Sect. II for
the scalar field, indicates that in order to determine their general
solution we must specify $\psi_+$ on $x^+=-L$ and $\psi_-$ on
$x^-=-L$.  The problem is thus to find expressions for the quantum
fields on these surfaces that furnish a representation of the
free-field canonical anticommutation relations, are properly causal,
etc.  The relevant commutators are
\bea
\left\{\psi_+(-L,x^-),\psi_+^\dagger(-L,y^-)\right\} &=& \d(x^--y^-)\\ 
\left\{\psi_-(x^+,-L),\psi_-^\dagger(x^+,-L)\right\} &=& \d(x^+-y^+)\\
\left\{\psi_+(-L,x^-),\psi_-^\dagger(-L,y^-)\right\} &=&
-{im\over4}\e(x^--y^-)\; .
\label{mixed}
\eea
In addition, all field operators should anticommute for space-like 
separations.

For $m=0$ the necessary construction has been given in Ref. \cite{7}.
After imposing antiperiodicity at equal time,
\be
\psi_\pm(-L,L) = -\psi_\pm(L,-L)\; ,
\ee
an argument identical to that given for the massless scalar shows that
$\psi_+$ may be chosen to be antiperiodic on $x^+=-L$ and $\psi_-$ may
be chosen to be antiperiodic on $x^-=-L$.  Thus we may write
$\psi_+(-L,x^-) = \psi(x^-)$ and $\psi_-(x^+,-L) = 
\widetilde{\psi}(x^+)$, where
\bea
\psi(x^-) &=& {1\over\sqrt{2L}}\sum_{n=1,3,\dots}^\infty
\left( b_n e^{-{i\over2}k^+_nx^-}
+d^\dagger_n e^{{i\over2}k^+_nx^-}\right) \label{psiexp}\\
\nonumber\\
\widetilde{\psi}(x^+) &=& {1\over\sqrt{2L}}\sum_{n=1,3,\dots}^\infty
\left( \b_n e^{-{i\over2}k^-_nx^+}
+\d^\dagger_n e^{{i\over2}k^-_nx^+}\right)\; ,
\label{psitexp}
\eea
and $k^\pm_n=n\pi/L$.  The correct commutation relations are
\bea
\left\{\psi(x^-),\psi^\dagger(y^-)\right\} &=&
\d(x^--y^-)\label{comm1}\\
\left\{\psit(x^+),\psit^\dagger(y^+)\right\} &=&
\d(x^+-y^+)\; ,\label{comm2}
\eea
with all other anticommutators vanishing, which correspond to
\be
\{b_n,b^\dagger_m\} =
\{d_n,d^\dagger_m\} =
\{\b_n,\b^\dagger_m\} =
\{\d_n,\d^\dagger_m\} = \d_{n,m}\; .
\label{fockcomms}
\ee
The Poincar\'e generators receive contributions from both wings of the
initial-value surface [see Eqn. (\ref{pmusym})], and correctly
translate all fields in $x^\pm$.  Note that if the degrees of freedom
represented by $\b_n$ and $\d_n$ are not included, then the theory
contains only right-moving particles and is not equivalent to the
equal-time theory of a massless Dirac fermion.

There are various ways of arriving at a correct construction for the
case $m\neq0$.  Here I shall simply motivate one possible way, and
then show that it works.

Let us start out by trying the same procedure as in the massless case.
That is, we choose $\psi_+$ to be an antiperiodic function $\psi(x^-)$
on $x^+=-L$ and $\psi_-$ to be an antiperiodic function $\psit(x^+)$
on $x^-=-L$.  We shall also attempt to preserve the simple commutation
relations (\ref{fockcomms}).  Now the Dirac equation allows
us to obtain $\psi_-$ on the surface $x^+=-L$:
\bea
\psi_-(-L,x^-) &=& \psi_-(-L,-L) 
-{im\over2}\int_{-L}^{x^-} dy^- \psi_+(-L,y^-)\\
&=& \widetilde{\psi}(-L)
-{im\over2}\int_{-L}^{x^-} dy^- \psi(y^-)
\eea
(the treatment of
$\psi_+(x^+,-L)$ is completely symmetric, of course).  It is helpful
to rewrite this in the form
\be
\psi_-(-L,x^-)=\psit(-L)
-{im\over4}\int_{-L}^L dy^- \e(x^--y^-)\psi(y^-)
-{im\over4}\int_{-L}^L dy^- \psi(y^-)\; ,
\label{naive}
\ee
from which we see that the simple commutation relations
(\ref{comm1})--(\ref{comm2}) do not correctly reproduce the
anticommutator (\ref{mixed}).  The second term on the RHS of
Eqn. (\ref{naive}) gives us what we want; the last term is the
troublemaker.  Let us attempt to fix this by defining $\psi_-(-L,x^-)$
to be
\be
\psi_-(-L,x^-)=\psit(-L)
-{im\over4}\int_{-L}^L dy^- \e(x^--y^-)\psi(y^-)\; ,
\ee
which gives the right commutator but changes the values of $\psi_-$ at
the endpoints.  We now have
\be
\psi_-(-L,\pm L) = \psit(-L) 
\mp {im\over4}\int_{-L}^L dy^- \psi(y^-)\; .
\label{pmbdy1}
\ee
Note that equal-time antiperiodicity then implies
\bea
\psi_-(L,-L) &=& -\psi_-(-L,L)\nonumber\\
&=& \psit(L) + {im\over4}\int_{-L}^L dy^- \psi(y^-)\; ,
\label{pmbdy2}
\eea
suggesting that we should also modify the value of $\psi_-$ at the
point $(L,-L)$.

We are thus led to seek an expression for $\psi_-(x^+,-L)$ that is
more complicated than a simple antiperiodic function $\psit$.
Furthermore, the modified $\psi_-(x^+,-L)$ should involve $\psi$, that
is, should depend on $\psi_+(-L,x^-)$.  A clue as to how to achieve
this is obtained by thinking about causality.  Recall that
$\psi_-(x^+,-L)$ and $\psi_+(-L,x^-)$ should anticommute for $x^+\neq
-L$ and $x^-\neq -L$.  It would therefore be dangerous to include in
$\psi_-(x^+,-L)$ anything involving $\psi$ except at the endpoints
$x^+=\pm L$; anything else would be likely to cause $\psi_-$ and
$\psi_+$ to fail to anticommute at space-like separation.  This
suggests a modification of $\psi_-(x^+,-L)$ at the endpoints only,
of the particular form
\be
\psi_-(x^+,-L)=\psit(x^+)
+{im\over4} \left[2+\e(x^+-L)-\e(x^++L)\right]
\int_{-L}^L dy^- \psi(y^+)\; .
\label{finalpm}
\ee
That is, we add in a discontinuity at each endpoint designed to
reproduce Eqns. (\ref{pmbdy1})--(\ref{pmbdy2}).  This leaves the
fields properly causal; away from the corner points
$\psi_-(x^+,-L)=\psit(x^+)$ and $\psi_+(-L,x^-)=\psi(x^-)$, which
anticommute.  Of course, the situation is symmetric with respect to
$+\leftrightarrow-$ so that we should also take
\be
\psi_+(-L,x^-)=\psi(x^-)+{im\over4}
\left[2+\e(x^--L)-\e(x^-+L)\right]\int_{-L}^L dy^+ 
\widetilde{\psi}(y^+)\; .
\label{finalpp}
\ee
Eqns. (\ref{finalpm}) and (\ref{finalpp}) are the final expressions
for the independent fields on their respective initial-value surfaces.

As mentioned previously, there are other ways of arriving at these,
but this one highlights the physical motivations behind the
construction.  Note that the fields are not strictly antiperiodic on
the initial surfaces.  That this must be the case follows immediately
from integrating the equation of motion and imposing equal-time
antiperiodicity; we obtain, e.g.,
\be
\psi_-(L,-L) + \psi_-(-L,-L) = 
{im\over2}\int_{-L}^L dx^- \psi_+(-L,x^-)\; .
\ee
We have constructed the fields to be as antiperiodic as possible,
however, while preserving the simple anticommutation relations
(\ref{comm1})--(\ref{comm2}).  The motivation for this is that
ultimately we would like to use this construction to obtain an
effective Hamiltonian for the usual light-cone degrees of freedom
($\psi_+$).  We would therefore like the formalism to resemble as much
as possible the usual one, with an antiperiodic $\psi_+$.  We could
perhaps have arrived at simpler expressions for the fields had we not
insisted on maintaining simple Fock space commutation relations.

The energy-momentum tensor has components
\be
T^{+-} = T^{-+} 
= m\left(\psi^\dagger_-\psi_+ + \psi^\dagger_+\psi_-\right)
\ee
\be
T^{++} = 2i\psi_+^\dagger\del_-\psi_+ + {\rm H.c.}
\label{tppf}
\ee
\be
T^{--} = 2i\psi_-^\dagger\del_+\psi_- + {\rm H.c.}
\label{tmmf}
\ee
and the Poincar\'e generators work out to be [Eqn. (\ref{pmusym})]
\bea
P^- &=& -{im^2\over4}\int_{-L}^L dx^- \int_{-L}^L dy^-
\psi^\dagger(x^-)\e(x^--y^-)\psi(y^-)
+2i\int_{-L}^L dx^+ \widetilde{\psi}^\dagger(x^+)\del_+
\widetilde{\psi}(x^+)
\nonumber\\ \nonumber \\
& &\qquad+m\widetilde{\psi}^\dagger(-L)\int_{-L}^L dx^-\psi(x^-)
+m\left[\int_{-L}^L dx^-\psi^\dagger(x^-)\right]\widetilde{\psi}(-L)
\eea
\bea
P^+ &=& -{im^2\over4}\int_{-L}^L dx^+ \int_{-L}^L dy^+
\widetilde{\psi}^\dagger(x^+)\e(x^+-y^+)\widetilde{\psi}(y^+)
+2i\int_{-L}^L dx^- \psi^\dagger(x^-)\del_-
\psi(x^-)
\nonumber\\ \nonumber \\
& &\qquad+m\psi^\dagger(-L)\int_{-L}^L dx^+\widetilde{\psi}(x^+)
+m\left[\int_{-L}^L dx^+\widetilde{\psi}^\dagger(x^+)\right]
\psi(-L)\; .
\eea
These can be shown to correctly translate all fields in $x^\pm$,
including the fields at the corner points, and furthermore satisfy
$[P^+,P^-]=0$.  We thus have a satisfactory representation of the
theory with the boundary degrees of freedom present.  Note the
presence of terms coupling together the fields on the two wings.

A few technical points are worth noting.  First, the discontinuities
in the fields at the corner points lead to singularities in their {\em
derivatives}, and these give rise to important contributions in the
pieces of $P^\mu$ involving $T^{++}$ and $T^{--}$.  Second, in
verifying the Heisenberg relations or the Poincar\'e algebra, it
should be kept in mind that since $\psi(x^-)$ is antiperiodic,
it follows that, for example,
\be
\left\{\psi(x^-),\psi^\dagger(-L)\right\} = \d(x^-+L) - \d(x^--L)\; .
\ee
Furthermore, when the argument of a delta function vanishes at an {\em
endpoint} of the integration region, the proper definition of the
integral is, e.g.,
\be
\int_{-L}^L dx^- f(x^-) \d(x^--L) = \ha f(L)\; .
\ee

It is now fairly simple to obtain suitable expressions for the fields
for use with the asymmetric initial-value surface shown in Fig. 2(b).
These may be obtained from Eqns. (\ref{finalpm}) and (\ref{finalpp})
by a shift of $x^+$ and a slight reorganization of the resulting
expressions \cite{11}.  For the independent fields we obtain
\be
\psi_+(0,x^-) = \psi(x^-)+{im\over4}
\left[2+\e(x^--L)-\e(x^-+L)\right]\int_{-L}^L dy^+ \e(y^+)
\widetilde{\psi}(y^+)
\ee
and
\be
\psi_-(x^+,\pm L) = \widetilde{\psi}(x^+)-{im\over4}
\left[\e(x^+)\pm1\right]\int_{-L}^L dy^- \psi(y^+)\; .
\ee
Integrating the equation of motion then gives us the constrained
fields:
\be
\psi_+(x^+,\pm L)=\psi(\pm L)-{im\over4}\int_{-L}^Ldy^+
\e(x^+-y^+)\widetilde{\psi}(y^+)
\ee
\be
\psi_-(0,x^-)=\widetilde{\psi}(0)-{im\over4}\int_{-L}^Ldy^-
\e(x^--y^-)\psi(y^-)\; .
\ee

The Poincar\'e generators are now obtained by integrating
\be
P^\pm=\ha\int_{-L}^0dx^+T^{-\pm}(x^+,L)
     +\ha\int_{-L}^Ldx^-T^{+\pm}(0,x^-)
     +\ha\int_0^Ldx^+T^{-\pm}(x^+,-L)\; .
\ee
We find
\bea
P^+ &=& -{im^2\over4}\int_{-L}^L dx^+ \int_{-L}^L dy^+
\widetilde{\psi}^\dagger(x^+)\e(x^+-y^+)\widetilde{\psi}(y^+)
+2i\int_{-L}^L dx^- \psi^\dagger(x^-)\del_-
\psi(x^-)
\nonumber\\ \nonumber \\
& &\qquad+m\psi^\dagger(-L)\int_{-L}^L dx^+\e(x^+)
\widetilde{\psi}(x^+)
+m\left[\int_{-L}^L dx^+\e(x^+)\widetilde{\psi}^\dagger(x^+)\right]
\psi(-L)
\eea
\bea
P^- &=& -{im^2\over4}\int_{-L}^L dx^- \int_{-L}^L dy^-
\psi^\dagger(x^-)\e(x^--y^-)\psi(y^-)
+2i\int_{-L}^L dx^+ \widetilde{\psi}^\dagger(x^+)\del_+
\widetilde{\psi}(x^+)
\nonumber\\ \nonumber \\
& &\qquad+m\widetilde{\psi}^\dagger(0)\int_{-L}^L dx^-\psi(x^-)
+m\left[\int_{-L}^L dx^-\psi^\dagger(x^-)\right]\widetilde{\psi}(0)\; .
\eea
Again, these can be shown to properly translate the fields and to
commute with each other.

\section{DISCUSSION}

I have argued that in a finite volume, introduced as an infrared
regulator, including degrees of freedom not present in the usual
light-cone formalism is necessary if the resulting theory is to be
equivalent to the corresponding equal-time theory.  This follows from
analyzing the equations of motion, as well as from constructing
Poincar\'e generators and other charges that are the same as the
operators we would construct in equal-time quantization.  I have also
attempted to give some flavor of the formalism necessary to achieve
this equivalence.  A major advantage of this approach is that it gives
a regulated theory but does not force the physical vacuum to be
trivial.  This is a welcome feature, as the QCD vacuum, for example,
is certainly not trivial.

Its main disadvantage is that it is quite complicated, even at the
level of free field theory.  In fact, it is more complicated than
equal-time field theory, so that if the point were to use it directly
as a calculational tool it would be largely useless.  However, recall
that vacuum triviality on the light cone really means that the effects
of the vacuum should be expressible in terms of local effective
interactions for a set of fields initialized at a fixed $x^+$.  That
is, we should be able to construct an effective theory which has a
trivial vacuum {\em state}, but nontrivial vacuum physics.  The
light-cone framework is the only one in which such an approach is
possible, due to the unusual connection between long distances (in
$x^-$) and high (light-cone) energies reflected in the dispersion
relation (\ref{disprel}).

\begin{figure}
\centerline{
\psfig{figure=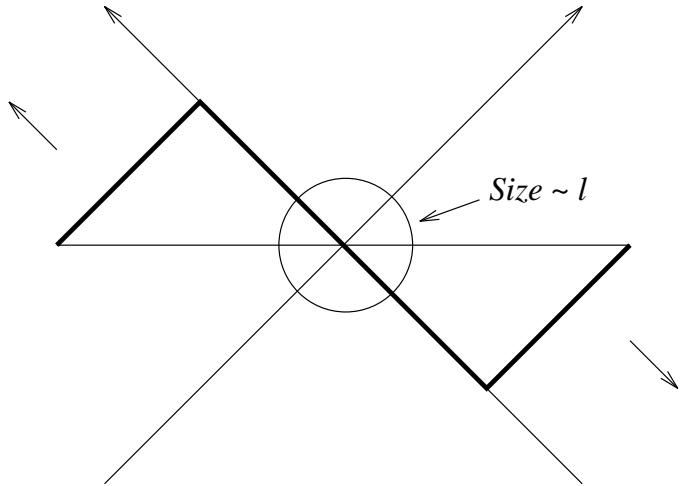}
}
\vspace{.1in}
\caption{Decoupling the boundaries in the large-$L$ limit.}
\end{figure}

This construction can be used to obtain such effective Hamiltonians in
a natural way.  The basic idea is to consider the limit in which the
box size $L$ becomes large compared to some fixed scale $l$ of
physical interest [Fig. 4].  In this case, the fields that live on the
boundaries appear to decouple from those that live at $x^+=0$; by
causality, quanta from the boundaries that can influence the region of
interest have their world lines pushed closer and closer to the light
cone as the box gets large.  This decoupling may be illusory, however,
because the couplings of these states can simultaneously diverge.  The
goal is to extract the finite remainders in this limit, and build them
in to an effective Hamiltonian for use an a ``traditional'' light-cone
calculation.  Details of some simple examples are given in \cite{11}.

This formulation of the light-cone theory with boundaries is also
instructive from the point of view of the puzzles listed in Sect. I.
For example, in the language of partial differential equations the
construction corresponds to the ``characteristic'' initial-value
problem, as distinct from the Cauchy problem.  This is perfectly well
defined from a mathematical point of view.  There is no real problem
with using surfaces of characteristics to specify initial conditions,
but more than one such surface is necessary in general.

The question of what boundary conditions may be imposed on the
characteristic surfaces can be addressed naturally in the course of
setting up a concrete realization of the scheme.  In general, imposing
strict periodicity or antiperiodicity is seen to be in conflict with
the dynamics, and results in the removal of degrees of freedom.

The massless limit of the construction is perfectly smooth; the
degrees of freedom that seem to appear only for $m=0$ are actually
necessary for the massive theory as well, and so are present from the
outset.  As discussed above, the confusion over whether the extra
fields are necessary in the massive case arises from considering the
artificially simple case of free theory, where the boundaries do
decouple for massive fields.  In general they {\em are} required, and
when they are present there is nothing discontinuous, or otherwise
unusual, about the massless limit.

Regarding the ``No-problem'' problem, the construction allows us to
study directly the effects of discarding various pieces of physics.
It would be very helpful to have a quantitative understanding, for
example, of what could be computed reliably without including the
boundaries.  The present scheme allows this to be studied concretely.
There are also more specific questions along these lines that can be
considered.  For example, what is the relation between the effective
interactions induced by the constrained zero modes in DLCQ and the
effective interactions one obtains in the full theory?  Do they in
fact capture some sort of ``mean field'' vacuum properties, and if so,
what can be calculated reliably in that framework without addressing
the full boundary problem?

It is perhaps worth restating that the potential advantage of
light-cone quantization over equal-time quantization is not that the
vacuum becomes trivial, but rather that the vacuum can trivially be
made trivial, by means of a simple cutoff on longitudinal momenta.
This has several immediate advantages.  When coupled with the insight
that the states thus removed are mainly high-energy states in 3+1
dimensions, it leads to the hope that we might be able to derive a
constituent {\em approximation} to QCD.  In fact, it is the only
framework that offers any realistic hope of achieving this.  In
addition, for problems where the vacuum does not play an important
physical role the ability to ignore it so simply may be quite useful.
In QCD, for example, it is presumably not necessary to completely
understand the pion or to have strictly linear long-range potentials
in order to obtain a reasonable description of, e.g., heavy quarkonia.
For this sort of problem it may be quite reasonable to ignore the
effective interactions that represent the vacuum, or, equivalently, to
use the conventional light-cone formalism without boundary fields.

For problems where the vacuum {\em does} play a role, however, its
structure must be properly accounted for.  The formalism described
here, or something equivalent, will be necessary to obtain the
effective interactions that mediate vacuum physics.  For QCD this can
be expected to be quite difficult, in part because the problem is
intrinsically nonperturbative.  Of course, the QCD vacuum is a
difficult problem in any (known) calculational scheme!  In any case,
the possibility of obtaining a constituent approximation to QCD,
complementary to the lattice, and of making contact with the natural
and very successful phenomenology based on the light-cone Fock
representation, makes this set of problems of considerable interest.

\acknowledgements

It is a pleasure to thank the organizers of {\sc Orbis Scientiae 1996}
for providing such a stimulating and enjoyable atmosphere.  This work
was done in collaboration with Kent Hornbostel, and I am grateful to
him for many enjoyable and illuminating discussions on the subject of
light-cone field theory.  I also thank Gary McCartor and Robert Perry
for many helpful conversations regarding the light-cone vacuum.  This
work was supported in part by a grant from the the U.S. Department of
Energy.


\begin{references}

\bibitem{1}
For an overview of QCD phenomenology from the light-cone point of
view, see:
S. J. Brodsky, these proceedings; 
S. J. Brodsky and D. G. Robertson, ``Light-Cone Quantization and QCD
Phenomenology,'' {\tt hep-ph/9511374}, to appear in the proceedings of
the ELFE Summer School and Workshop on Confinement Physics, Cambridge,
UK, July 1995.

\bibitem{2}
K. Hornbostel, Phys. Rev. D {\bf 45}, 3781 (1992).

\bibitem{3}
R. J. Perry, Ann. Phys. {\bf 232}, 116 (1994);
K. G. Wilson, T. S. Walhout, A. Harindranath, W.-M. Zhang,
R. J. Perry, and St.~D.~Glazek, Phys. Rev. D {\bf 49}, 6720 (1994).

\bibitem{4}
S. J. Brodsky and H.-C. Pauli, in {\it Recent Aspects of Quantum
Fields}, H.~Mitter and H.~Gausterer, Eds., Lecture Notes in Physics,
Vol. 396 (Springer-Verlag, 1991), and references therein.

\bibitem{5}
T. Heinzl, S. Krusche, S. Simb\"urger, and E. Werner, Z. Phys. C {\bf
56}, 415 (1992);
D. G. Robertson, Phys. Rev. D {\bf 47}, 2549 (1993);
C. M. Bender, S. S. Pinsky, and B. van de Sande, Phys. Rev. D {\bf
48}, 816 (1993);
S. S. Pinsky and B. van de Sande, Phys. Rev. D {\bf 49}, 2001 (1994);
J. Hiller, S. S. Pinsky, and B. van de Sande, Phys. Rev. D {\bf 51},
726 (1995).  

\bibitem{6}
F. Rohrlich, Acta Phys. Austr. {\bf 32}, 87 (1970);
P. J. Steinhardt, Ann. Phys. {\bf 128}, 425 (1980);
T. Heinzl and E. Werner, Z. Phys. C {\bf 62}, 521 (1994). 

\bibitem{7}
G. McCartor, Z. Phys. C {\bf 41}, 271 (1988). 

\bibitem{8}
H. Bergknoff, Nucl. Phys. B {\bf 122}, 215 (1977);
T. Eller, H.-C. Pauli, and S. J. Brodsky, Phys. Rev. D {\bf 35}, 1493
(1987).

\bibitem{9}
G. McCartor, Z. Phys. C {\bf 52}, 611 (1991); {\it ibid. \bf 64}, 349
(1994).

\bibitem{10}
K. Hornbostel, Ph. D. thesis, Stanford University (1988).

\bibitem{11}
K. Hornbostel and D. G. Robertson, manuscript in preparation (1996).

\end{references}
\end{document}